\documentclass[12pt, draftclsnofoot, onecolumn]{IEEEtran}
\usepackage{amsmath,amssymb,stmaryrd}
\usepackage[a4paper,top=2cm,bottom=2cm,left=2cm,right=2cm]{geometry}
\usepackage[english]{babel}
\usepackage{cases}
\usepackage{setspace}
\usepackage{pifont}
\usepackage{cite}
\usepackage[boxed]{algorithm2e}
\usepackage{graphicx}
\usepackage{epstopdf}
\usepackage[justification=centering]{caption}
\newtheorem{theorem}{Theorem}[section]
\newtheorem{lemma}{Lemma}[section]

\title{Resource allocation for reconfigurable intelligent surface aided broadcast channels}
\author{Cong Sun,~\IEEEmembership{Member,~IEEE,} Xian Liu, Bile Peng,~\IEEEmembership{Member,~IEEE,} and Eduard Jorswieck,~\IEEEmembership{Fellow,~IEEE}
\thanks{C. Sun and X. Liu are with Beijing University of Posts and Telecommunications, Beijing, China, 100876 (e-mail: \{suncong86, liuxian\}@bupt.edu.cn).}
\thanks{B. Peng and E. Jorswieck are with Institute for Communications Technology, TU Braunschweig, 38106 Braunschweig, Germany (e-mail: \{peng, jorswieck\}@ifn.ing.tu-bs.de).} }

\begin{document}
\maketitle

\begin{abstract}
A two-user downlink network aided by a reconfigurable intelligent surface is considered. The weighted sum signal to interference plus noise ratio maximization and the sum rate maximization models are presented, where the precoding vectors and the RIS matrix are jointly optimized. Since the optimization problem is non-convex and difficult, new approximation models are proposed.
The upper bounds of the corresponding objective functions are derived and maximized. Two new algorithms based on the alternating direction method of multiplier are proposed. It is proved that the proposed algorithms converge to the KKT points of the approximation models as long as the iteration points converge. Simulation results show the good performances of the proposed models compared to state of the art algorithms.

\end{abstract}

\section{Introduction}
Reconfigurable Intelligent Surface (RIS) is a kind of electromagnetic metasurface, which passively reflects microwave signals with tunable phase shifts \cite{Tang,Basar,Wu1}. Recently, RIS technique becomes popular in the new generation of wireless communications. Acting like passive relay, RIS assists communications and provides multi-path gains. RIS technique is different from relay in the following aspects \cite{Renzo,Bjornson,Sun2016}: digital signal processing and power amplification are required at relay for all relaying nodes; on the contrast, RIS reflects signals in passive mode, which changes the signals' phase shifts, and thus has less power consumption compared to relay.

There are many works concentrating on the fundamental limits of the RIS aided communications, from point-to-point to Multiple-Input Multiple-Output (MIMO) networks. In \cite{Zappone}, the authors explore the optimal number of RIS elements for maximizing the rate and the energy efficiency of the point-to-point channel aided by RIS without direct link. The unimodal function property is exploited. More researches focus on Multiple-Input Single-Output (MISO) downlink channels. For single user links, the rate maximization model is considered in \cite{Wu2018}, and both centralized and distributed algorithms are proposed. With statistical channel state information (CSI), \cite{Hu} maximizes the ergodic rate, where both Rician fading and Rayleigh fading scenarios are considered. However, the single user case cannot be directly extended to multiple user case. Not only interference is introduced due to multiple users, but also different measurements of Quality of Service (QoS) are supported. The power minimization models with SINR constraints for both single and multiple users are considered in \cite{Wu}, where the alternating optimization (AO) and semi-definite relaxation (SDR) techniques are applied. In \cite{Li}, the power minimization and admission control are considered in a single optimization model. The authors simplify the complicated problem by the difference-of-convex and linear approximation techniques. Considering the minimum Signal-to-Interference-plus-Noise (SINR) maximization problem, \cite{Nadeem} applies the approximated optimal linear precoding, and uses the projected gradient method to solve the subproblem for the RIS parameters, where the rank-one and high rank channel coefficient matrix cases are discussed. The authors in \cite{Gao} jointly optimize the precoding beamforming vectors and the RIS matrix, to maximize the sum rate with power and proportional rate constraints. The precoding vectors are chosen according to the zero forcing (ZF) technique, and the power allocation and the RIS matrix are solved by the dual methods alternatively. The weighted sum rate maximization problem is discussed in \cite{Guo}, where the sum fractional objective function is reformulated via the fractional optimization technique. The block coordinate descent (BCD) method is applied, and the subproblem for RIS parameters are solved by three different algorithms using projected gradient and alternating direction method of multiplier (ADMM) techniques. ADMM method is also applied for the weighted sum SINR maximization problem \cite{Liu}, where the authors approximate the SINRs by their achievable upper bounds. Another mean square error minimization model is presented in \cite{Rehman}, and the nonconvex problem is solved via AO, BCD and vector approximate message passing techniques. In \cite{Peng}, the fully conventionally network and the weighted minimum mean square error techniques are combined together for the weighted sum rate maximization problem.

Recently, RIS aided MIMO networks are taken into consideration \cite{Dai}. For single user MIMO networks, \cite{Chang} considers the rate maximization problem with transmit power control and RIS phase shift constraints. The authors apply the augmented Lagrangian function technique to transform the constrained optimization into an unconstrained problem, and solve it by the quasi-Newton method. Considering the same model, \cite{Zhang} analyzes the power allocation schemes in different Signal-to-Noise-Ratio (SNR) scenarios. For low SNR regime, transmit power concentrates on single stream; for high SNR regime, every stream is distributed with equal power. The work \cite{Pan} proposes a multi-cell MIMO model aided by one piece of RIS, and maximizes the weighted sum rate through alternatively optimizing precoding matrices and RIS phase shift parameters.

Besides resource allocation models, researchers also extend the RIS technique to various other communication scenarios like energy efficiency \cite{Huang}, cognitive radio \cite{Yuan}, energy harvesting \cite{Pan1}, NOMA \cite{Mu,Jiang}, mmWave \cite{Jorswieck} and wiretap channel \cite{Chu}.


In most works, precoding vectors and RIS matrix are jointly designed through the AO technique. The iterations might converge to a point without any theoretical guarantee. In this paper, we will use the ZF idea and the Maximum Ratio Transmission (MRT) precoding technique to reformulate the optimization model, which only optimizes the RIS parameters. This will not only save computational cost, but also avoid the disadvantage of AO.

In this work, the two-user downlink channel aided by RIS is considered. The system model is presented in Section 2. In Section 3, the weighted SINR maximization model and the corresponding algorithm are proposed. Section 4 mainly discusses the sum rate maximization model as well as algorithm. Simulation results in Section 5 show the effectiveness of the proposed models and methods. Conclusions are shown in Section 6.

\emph{Notation}: $\mathbb{R}$ and $\mathbb{C}$ represent the real domain and the complex domain, respectively. $(\cdot)^{T}$ and $(\cdot)^{H}$ mean the transpose and the Hermitian, respectively. $\bar{a}$ is the conjugate of the complex number $a$. $\mathbf{I}_{N}$ represents the $N\times N$ identity matrix. $\textrm{Diag}(\mathbf{a})$ represents a diagonal matrix with diagonal elements as the elements of the vector $\mathbf{a}$. $\mathbf{A}^{+}$ and $\textrm{rank}(\mathbf{A})$ represent the pseudo inverse and the rank of any matrix $\mathbf{A}$, respectively. $\odot$ represents for the pointwise product. $\lambda_{\textrm{max}}(\mathbf{B})$ is the maximum eigenvalue of the square matrix $\mathbf{B}$. $\textrm{Re}(\cdot)$ denotes the real part of a complex number. $(\cdot)_{R}$ and $(\cdot)_{I}$ mean the real and imaginary parts of a matrix or vector, respectively. $\mathbb{E}(\cdot)$ denotes the statistical expectation. $\mathcal{CN}(\mu,\sigma^{2})$ represents the complex Gaussian distribution with mean as $\mu$ and variance as $\sigma^{2}$.

\section{System model}
\begin{figure}
  \begin{minipage}{1.0\linewidth}
  \centering
  \centerline{\includegraphics[width=10cm]{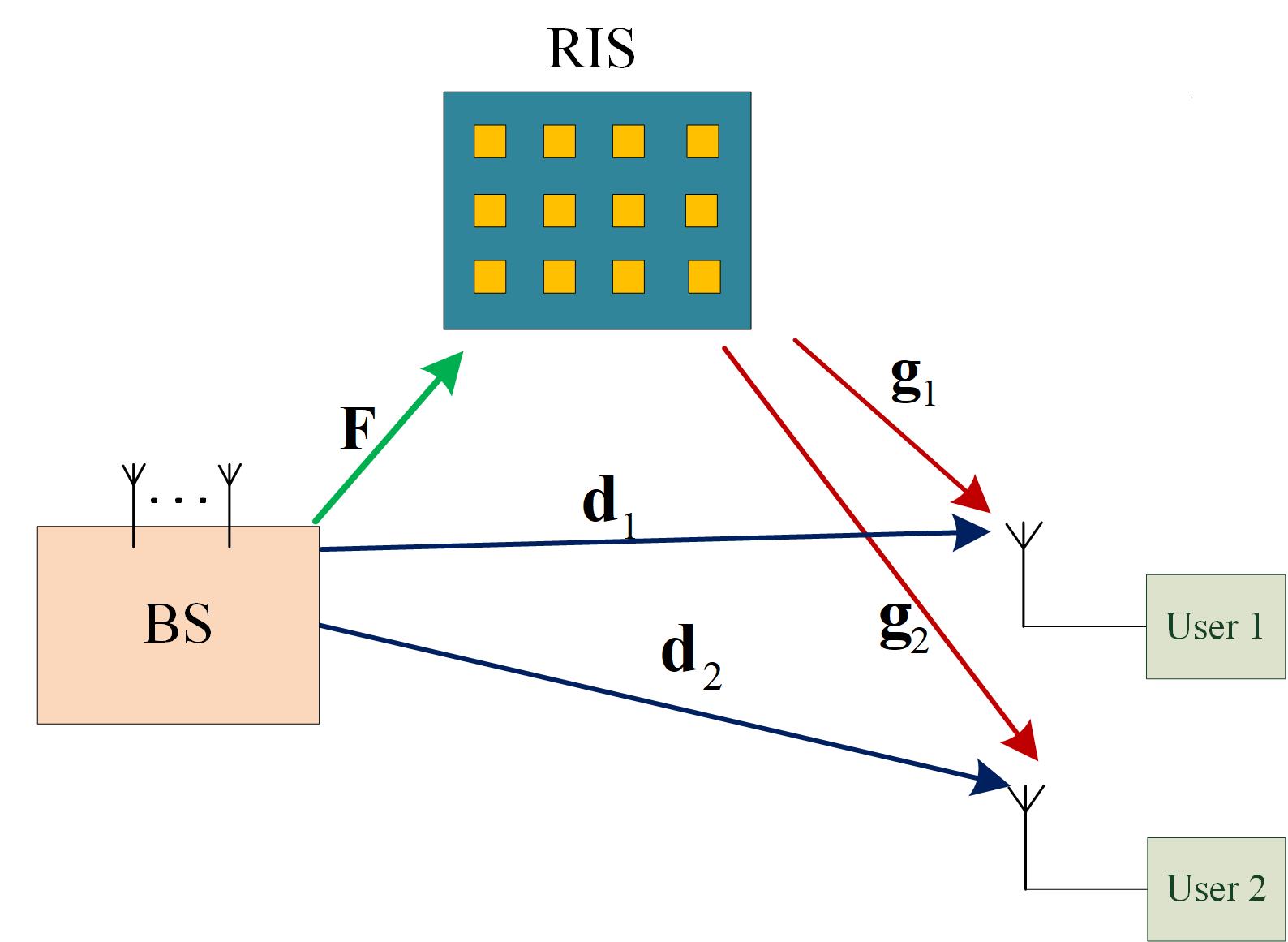}}
  \caption{RIS-aided downlink network with two users}
  \label{fig1}
  \end{minipage}
\end{figure}
Consider a two-user downlink network as in Fig. \ref{fig1}. The communications between $M$-antenna base station (BS) and the two single-antenna users are assisted by an RIS equipped with $N$ passive reflecting units. Usually $N\gg M$. Assume that $\mathbf{F}\in \mathbb{C}^{N\times M}$ is the channel matrix between the BS and the RIS; $\mathbf{d}_j\in \mathbb{C}^{M\times 1}$ and $\mathbf{g}_j\in \mathbb{C}^{N\times 1}$ are the channel vectors from the BS and the RIS to user $j$, respectively, $j=1,2$. Here we assume that all the channel state information (CSI) is perfectly known through pilot based channel estimations \cite{You}.

The transmission process is as follows. The BS broadcasts the signal:
$$\mathbf{x}_{\textrm{BS}}=\sum_{j=1}^2 \mathbf{w}_js_j,$$
where $\mathbf{w}_j\in \mathbb{C}^{M\times 1}$ is the precoding vector, and $s_j$ satisfying $\mathbb{E}(|s_j|^{2})=1$ denotes the transmit signal to user $j$. The signal $\mathbf{x}_{\textrm{BS}}$ is transmitted to each user via two paths. The first path is the direct link from the BS to the two users, where user $j$ receives $\mathbf{d}_j^{H}\mathbf{x}_{\textrm{BS}}$, $j=1, 2$. The second path is from the BS to user $j$ through the RIS. The RIS first receives $\mathbf{Fx}_{\textrm{BS}}$ and then reflects it to user $j$ as $\boldsymbol{\Theta} \mathbf{Fx}_{\textrm{BS}}$, where $\boldsymbol{\Theta}=\textrm{Diag}(\theta_{1},\ldots,\theta_{N})$ is the RIS matrix and $\theta_{1},\ldots,\theta_{N}$ denote $N$ reflection coefficients. The received signal of user 1 is
\begin{eqnarray*}
y_1=(\mathbf{d}_{1}^{H}+\mathbf{g}_{1}^{H}\boldsymbol{\Theta} \mathbf{F})\mathbf{x}_{\textrm{BS}}+n_1
=\underbrace{(\mathbf{d}_{1}+\mathbf{F}^{H}\boldsymbol{\Theta}^{H}\mathbf{g}_{1})^{H}\mathbf{w}_{1}s_{1}}_{\textrm{desired signal}}+\underbrace{(\mathbf{d}_{1}+\mathbf{F}^{H}\boldsymbol{\Theta}^{H}\mathbf{g}_{1})^{H}\mathbf{w}_{2}s_{2}}
_{\textrm{interference}}+\underbrace{n_{1}}_{\textrm{noise}},
\end{eqnarray*}
where $n_{1}\sim\mathcal{CN}(0,\sigma_{1}^{2})$ is the local noise at user 1. Here we suppose all the signals and noises are independent of each other. The SINR of user 1 is given by
\begin{eqnarray} \label{eq:eq1}
\textrm{SINR}_{1}=\frac{|(\mathbf{d}_{1}+\mathbf{F}^{H}\boldsymbol{\Theta}^{H}\mathbf{g}_{1})^{H}\mathbf{w}_{1}|^{2}}{|(\mathbf{d}_{1}
+\mathbf{F}^{H}\boldsymbol{\Theta}^{H}\mathbf{g}_{1})^{H}\mathbf{w}_{2}|^{2}+\sigma_{1}^{2}}.
\end{eqnarray}
Similarly, we have the SINR of user 2 as
$$
\textrm{SINR}_2=\frac{|(\mathbf{d}_2+\mathbf{F}^{H}\boldsymbol{\Theta}^{H}\mathbf{g}_2)^{H}\mathbf{w}_2|^{2}}{|(\mathbf{d}_2
+\mathbf{F}^{H}\boldsymbol{\Theta}^{H}\mathbf{g}_2)^{H}\mathbf{w}_1|^{2}+\sigma_2^{2}}.
$$

In the following, we will focus on the model to maximize the users' QoS, while the BS transmit power budget is restricted:
\begin{eqnarray} \label{eq:power}
\|\mathbf{w}_1\|^2+\|\mathbf{w}_2\|^2\leq P_T.
\end{eqnarray}
Here the continuous phase shifts as well as amplitude adaptation are applied to the RIS parameters. That is,
$$
|\theta_{l}|\leq 1,l=1,\ldots,N.
$$
Two different measures of QoS will be considered in Section \ref{sec:wsinr} and \ref{sec:sr}.

\section{Weighted sum SINR maximization model}\label{sec:wsinr}
First, the SINR of each user is used as its QoS measurement. We would like to maximize the weighted sum SINRs under the transmit power and RIS constraints. The corresponding optimization problem is as follows:
\begin{subequations} \label{eq:wSINR}
\begin{eqnarray}
\hspace{-0.28cm}&\displaystyle\max_{\mathbf{w}_{1},\mathbf{w}_{2},\boldsymbol{\Theta}}& \lambda \textrm{SINR}_{1}(\mathbf{w}_{1},\mathbf{w}_{2},\boldsymbol{\Theta})
+(1-\lambda) \textrm{SINR}_{2}(\mathbf{w}_{1},\mathbf{w}_{2},\boldsymbol{\Theta})\label{eq:wSINR.1}\\
\hspace{-0.28cm}&\textrm{s.~t.}& \|\mathbf{w}_1\|^2+\|\mathbf{w}_2\|^2\leq P_T,\\
\hspace{-0.28cm}&&\boldsymbol{\Theta}=\textrm{Diag}(\theta_{1},\ldots,\theta_{N}),|\theta_{l}|\leq 1,l=1,\ldots,N.
\end{eqnarray}
\end{subequations}
Here $\lambda\in[0,1]$ is the weight. By solving problem (\ref{eq:wSINR}), we can calculate the convex hull of the system's Pareto boundary of the SINR region. Due to the sum ratio expression and the coupled variables in the objective function, problem (\ref{eq:wSINR}) is nonconvex, and is thus difficult to solve. Next, we derive a suitable substitute of (\ref{eq:wSINR}) and propose an efficient algorithm to solve the approximated problem.


\subsection{Problem approximation}\label{sec:3.2}
It is trivial to see that $\textrm{SINR}_j, j=1,2$ are upper bounded in the following way.
\begin{eqnarray} \label{eq:eq8}
\textrm{SINR}_j \leq \frac{|\mathbf{h}_j^{H}\mathbf{w}_j|^{2}}{\sigma_j^{2}}
\leq \frac{\omega_j^2P_T\|\mathbf{h}_j\|_{2}^{2}}{\sigma_j^{2}},
\end{eqnarray}
where $\mathbf{h}_j=\mathbf{d}_j+\mathbf{F}^{H}\boldsymbol{\Theta}^{H}\mathbf{g}_j$ is the equivalent channel from the BS to user $j$; $\omega_j^2P_T$ is the transmit power allocated to user $j$, and $\omega_j$ will be further explained in (\ref{eq:eq9}). The corresponding upper bound is achievable in the following way.\\
1. Select a proper RIS matrix $\boldsymbol{\Theta}$ such that the two equivalent channels are orthogonal to each other:
$$
\mathbf{h}_1^H\mathbf{h}_2=0.
$$
2. Apply the MRT precoding technique:
\begin{eqnarray} \label{eq:eq9}
\mathbf{w}_j=\omega_j\sqrt{P_T}\frac{\mathbf{h}_j}{\|\mathbf{h}_j\|_{2}}.
\end{eqnarray}
Here $\omega_j, j=1,2$ satisfying $\omega_1^2+\omega_2^2\leq1$ represents the power allocation scheme.

Next, we maximize the achievable upper bound of (\ref{eq:wSINR.1}) instead.
Let $\mathbf{x}=(\theta_{1},\ldots,\theta_{N})^{H}$, $\mathbf{G}_j=\textrm{Diag}(\mathbf{g}_j)$ and $\mathbf{F}_j=\mathbf{F}^{H}\mathbf{G}_j$. From the fact that $\mathbf{F}^{H}\boldsymbol{\Theta}^{H}\mathbf{g}_j=\mathbf{F}_j\mathbf{x}$, we have the following problem:
\begin{eqnarray}\label{eq:eq16}
&\displaystyle\max_{\mathbf{x}\in\mathbb{C}^{N\times 1}, \omega_1, \omega_2\in\mathbb{R}}& \lambda\frac{\omega_1^2P_T}{\sigma_1^2}\|\mathbf{d}_{1}+\mathbf{F}_{1}\mathbf{x}\|_{2}^{2}
+(1-\lambda)\frac{\omega_2^2P_T}{\sigma_2^2}\|\mathbf{d}_{2}+\mathbf{F}_{2}\mathbf{x}\|_{2}^{2}\nonumber\\
&\textrm{s.~t.}& \omega_1^2+\omega_2^2\leq1,\nonumber\\
&& (\mathbf{d}_{1}+\mathbf{F}_{1}\mathbf{x})^{H}(\mathbf{d}_{2}+\mathbf{F}_{2}\mathbf{x})=0, \nonumber\\
&& |x_{l}|\leq 1,l=1,2,\ldots,N.
\end{eqnarray}
The above problem is not well-defined. It is trivial to see that the optimal $\omega_1$ and $\omega_2$ should be either $1$ or $0$, which degenerates to the single user case. But we would like to consider the multi-user case. Thus equal transmit power is allocated to each user. Without loss of generality, we suppose $\sigma_1^2=\sigma_2^2=\sigma^2$. The corresponding optimization problem becomes:
\begin{subequations} \label{eq:eq4}
\begin{eqnarray}
&\displaystyle\max_{\mathbf{x}\in\mathbb{C}^{N\times 1}}& \lambda\|\mathbf{d}_{1}+\mathbf{F}_{1}\mathbf{x}\|_{2}^{2}+(1-\lambda)\|\mathbf{d}_{2}+\mathbf{F}_{2}\mathbf{x}\|_{2}^{2}\label{eq:eq4.0}\\
&\textrm{s.~t.}& (\mathbf{d}_{1}+\mathbf{F}_{1}\mathbf{x})^{H}(\mathbf{d}_{2}+\mathbf{F}_{2}\mathbf{x})=0, \label{eq:eq4.1}\\
&& |x_{l}|\leq 1,l=1,2,\ldots,N.\label{eq:eq4.2}
\end{eqnarray}
\end{subequations}
The following theorem shows the conditions which guarantee the feasibility of problem (\ref{eq:eq4}).
\begin{theorem} \label{theo:1}
If the conditions $\|\mathbf{F}_j^{+}\mathbf{d}_j\|_{\infty}\leq 1$ and $\textrm{rank}(\mathbf{F}_j)=M$ hold for $j=\{1,2\}$, then problem (\ref{eq:eq4}) is feasible.
\end{theorem}
The detailed proof is shown in Appendix \ref{ap:ap1}. It is worth noting that in practice the parameters usually meet the requirement of Theorem \ref{theo:1}. Consequently, problem (\ref{eq:eq4}) is feasible. Furthermore, any feasible point of problem (\ref{eq:eq4}) is feasible for problem (\ref{eq:wSINR}). Thus, the optimum of problem (\ref{eq:eq4}) provides a lower bound for the objective function of problem (\ref{eq:wSINR}).

in the following part, problem (\ref{eq:eq4}) is solved instead of problem (\ref{eq:wSINR}). Problem (\ref{eq:eq4}) can be treated as a QCQP problem, however with $N+1$ constraints. The classic SDR method could be used \cite{Luo}. Since the number of RIS reflection elements $N$ is usually large, the achieved relaxed solution might not be rank one. Then we need to apply the randomization technique to generate a feasible point of problem (\ref{eq:eq4}), which loses optimality guarantee and may even fail to find a feasible solution.

\subsection{ADMM based method} \label{sec:ADMM}
The main difficulty for solving problem (\ref{eq:eq4}) stems from the complicated constraints. (\ref{eq:eq4.1}) is a single constraint which is nonconvex, while (\ref{eq:eq4.2}) represents $N$ convex constraints. This means problem (\ref{eq:eq4}) has many constraints which form a nonconvex feasible region. Due to the above reason, we adopt the ADMM method to solve problem (\ref{eq:eq4}). We first introduce an auxiliary variable $\mathbf{y}$ and an extra equality constraint $\mathbf{y}=\mathbf{x}$, to separate two kinds of constraints (\ref{eq:eq4.1}) and (\ref{eq:eq4.2}):
\begin{eqnarray}
&\displaystyle\max_{\mathbf{x}, \mathbf{y}\in\mathbb{C}^{N\times 1}}&
\lambda\|\mathbf{d}_{1}+\mathbf{F}_{1}\mathbf{x}\|_{2}^{2}+(1-\lambda)\|\mathbf{d}_{2}+\mathbf{F}_{2}\mathbf{x}\|_{2}^{2}\nonumber\\
&\textrm{s.~t.}& (\mathbf{d}_{1}+\mathbf{F}_{1}\mathbf{x})^{H}(\mathbf{d}_{2}+\mathbf{F}_{2}\mathbf{x})=0, \nonumber\\
&& |y_{l}|\leq 1,l=1,2,\ldots,N, \nonumber\\
&& \mathbf{y}=\mathbf{x}.
\end{eqnarray}
Then the Augmented Lagrangian penalty function technique is applied, so that the constraint $\mathbf{y}=\mathbf{x}$ is penalized to the objective function \cite{yx}. The problem becomes:
\begin{eqnarray} \label{eq:eq5}
&\displaystyle\min_{\mathbf{x}, \mathbf{y}\in\mathbb{C}^{N\times 1}}& - [\lambda\|\mathbf{d}_{1}+\mathbf{F}_{1}\mathbf{x}\|_{2}^{2}+(1-\lambda)\|\mathbf{d}_{2}+\mathbf{F}_{2}\mathbf{x}\|_{2}^{2}]
+\textrm{Re}[\boldsymbol{\mu}^{H}(\mathbf{x}-\mathbf{y})]+\frac{\rho}{2}\|\mathbf{x}-\mathbf{y}\|_{2}^{2}\nonumber\\
&\textrm{s.~t.}& (\mathbf{d}_{1}+\mathbf{F}_{1}\mathbf{x})^{H}(\mathbf{d}_{2}+\mathbf{F}_{2}\mathbf{x})=0,\nonumber\\
&& |y_{l}|\leq 1,l=1,2,\ldots,N,
\end{eqnarray}
where $\boldsymbol{\mu}$ and $\rho$ are the corresponding Lagrange multiplier and penalty factor, respectively. Following the ADMM framework \cite{Eckstein}, we alternatively update the variables $\mathbf{x}$ and $\mathbf{y}$. And the Lagrange multiplier $\boldsymbol{\mu}$ and the penalty factor $\rho$ are updated accordingly.

\subsubsection{Subproblem for $\mathbf{x}$} \label{sec:x}
If we fix $\mathbf{y}$, the subproblem with respect to variable $\mathbf{x}$ is:
\begin{subequations}\label{eq:eq6}
\begin{eqnarray}
\hspace{-0.25cm}&\displaystyle\min_{\mathbf{x}\in\mathbb{C}^{N\times 1}}& f(\mathbf{x})=\mathbf{x}^{H}\mathbf{A}_{1}\mathbf{x}+\mathbf{x}^{H}\mathbf{a}_{1}+\mathbf{a}_{1}^{H}\mathbf{x} \label{eq:eq6.1}\\
\hspace{-0.25cm}&\textrm{s.~t.}& g(\mathbf{x})=\mathbf{x}^{H}\mathbf{A}_{2}\mathbf{x}+\mathbf{x}^{H}\mathbf{a}_{2}+\mathbf{a}_{3}^{H}\mathbf{x}+a_{4}=0,\label{eq:eq6.2}
\end{eqnarray}
\end{subequations}
where $\mathbf{A}_{1}=-\lambda\mathbf{F}_{1}^{H}\mathbf{F}_{1}
-(1-\lambda)\mathbf{F}_{2}^{H}\mathbf{F}_{2}+\frac{\rho}{2}\mathbf{I}_{N}$, $\mathbf{A}_{2}=\mathbf{F}_{1}^{H}\mathbf{F}_{2}$,
$\mathbf{a}_{1}=-\lambda\mathbf{F}_{1}^{H}\mathbf{d}_{1}
-(1-\lambda)\mathbf{F}_{2}^{H}\mathbf{d}_{2}-\frac{\rho}{2}\mathbf{y}+\frac{1}{2}\boldsymbol{\mu}$, \\ $\mathbf{a}_{2}=\mathbf{F}_{1}^{H}\mathbf{d}_{2}$, $\mathbf{a}_{3}=\mathbf{F}_{2}^{H}\mathbf{d}_{1}$ and $a_{4}=\mathbf{d}_{1}^{H}\mathbf{d}_{2}$.
Problem (\ref{eq:eq6}) is a QCQP problem with only one constraint. Let $\mathbf{F}_{12}=\lambda\mathbf{F}_{1}^{H}\mathbf{F}_{1}
+(1-\lambda)\mathbf{F}_{2}^{H}\mathbf{F}_{2}$. (\ref{eq:eq6}) is guaranteed to be bounded below when
\begin{eqnarray}\label{eq:eq76}
\rho>2\lambda_{\textrm{max}}(\mathbf{F}_{12}).
\end{eqnarray}
Then we just apply the SDR method to solve subproblem (\ref{eq:eq6}) and obtain its optimal solution.

Otherwise, (\ref{eq:eq6}) might be unbounded. In this case, we would like to find a new iteration point which is feasible and improves the objective function, such that the following theorem holds.
\begin{theorem}
Suppose $\mathbf{x}_k$ is the feasible iteration point from the $k$-th iteration. In the case that $\rho\leq 2\lambda_{\textrm{max}}(\mathbf{F}_{12})$, the new obtained iteration point $\mathbf{x}_{k+1}$
satisfies
$$
g(\mathbf{x}_{k+1})=0, f(\mathbf{x}_{k+1})<f(\mathbf{x}_{k}).
$$
\end{theorem}
\emph{Proof:} We prove the theorem by constructing $\mathbf{x}_{k+1}$. If (\ref{eq:eq6}) is unbounded, there must exist a direction $\mathbf{s}_k\in\mathbb{C}^{N}$, such that going along $\mathbf{s}_k$ the iteration point stays in the feasible region and the objective function value approaches $-\infty$. That is,
\begin{eqnarray} \label{eq:eq15}
g(\mathbf{x}_{k}+\alpha\mathbf{s}_k)=0, \forall \alpha\geq0; f(\mathbf{x}_{k}+\alpha\mathbf{s}_k)\xrightarrow{\alpha\rightarrow+\infty}-\infty.
\end{eqnarray}
According to (\ref{eq:eq15}), for $\forall \alpha\geq0$, we have
\begin{eqnarray}\label{eq:eq70}
&&g(\mathbf{x}_{k}+\alpha\mathbf{s}_k)\nonumber\\
&=& g(\mathbf{x}_{k})+[\mathbf{s}_k^{H}(\mathbf{F}_{1}^{H}\mathbf{d}_{2}+\mathbf{F}_{1}^{H}\mathbf{F}_{2}\mathbf{x}_{k})
+(\mathbf{F}_{2}^{H}\mathbf{d}_{1}+\mathbf{F}_{2}^{H}\mathbf{F}_{1}\mathbf{x}_{k})^{H}\mathbf{s}_k]\alpha
+(\mathbf{s}_k^{H}\mathbf{F}_{1}^{H}\mathbf{F}_{2}\mathbf{s}_k)\alpha^2\nonumber\\
&=&0.
\end{eqnarray}
Besides, the feasibility of $\mathbf{x}_k$ gives $g(\mathbf{x}_{k})=0$. Then we must have
\begin{subequations}\label{eq:eq7}
\begin{numcases}{}
 \mathbf{s}_k^{H}\mathbf{B}\mathbf{s}_k=0,\label{eq:eq7.1}\\[0.1cm]
 \mathbf{s}_k^{H}\mathbf{c}_{1}+\mathbf{c}_{2}^{H}\mathbf{s}_k=0,\label{eq:eq7.2}
 \end{numcases}
\end{subequations}
where $\mathbf{B}=\mathbf{F}_{1}^{H}\mathbf{F}_{2} \in \mathbb{C}^{N\times N}$, $\mathbf{c}_{1}=\mathbf{F}_{1}^{H}\mathbf{d}_{2}+\mathbf{F}_{1}^{H}\mathbf{F}_{2}\mathbf{x}_{k} \in \mathbb{C}^{N\times 1}$, $\mathbf{c}_{2}=\mathbf{F}_{2}^{H}\mathbf{d}_{1}+\mathbf{F}_{2}^{H}\mathbf{F}_{1}\mathbf{x}_{k} \in \mathbb{C}^{N\times 1}$. We would like to solve the nonlinear equation system (\ref{eq:eq7}) to obtain a non-zero $\mathbf{s}_k$. However it is difficult to deal with directly due to the quadratic equation (\ref{eq:eq7.1}). So we use the linear equation
\begin{eqnarray}\label{eq:eq10}
\mathbf{B}\mathbf{s}_k=\mathbf{0}
\end{eqnarray}
to replace (\ref{eq:eq7.1}). Equations (\ref{eq:eq10}) and (\ref{eq:eq7.2}) consist of a linear system. It is trivial to see that the solution of the new linear system must satisfy (\ref{eq:eq7}). Rewrite the linear system in real domain, and it becomes
\begin{eqnarray}\label{eq:eq14}
\mathbf{W}\tilde{\mathbf{s}}=\mathbf{0},
\end{eqnarray}
where $\mathbf{s}_k=\mathbf{s}_{1}^k+\mathbf{s}_{2}^ki$, $\mathbf{s}_{1}^k$ and $\mathbf{s}_{2}^k$ are real and imaginary parts of $\mathbf{s}_k$, respectively;
\begin{eqnarray*}
\mathbf{W}={\left( \begin{array}{ccc}
\mathbf{B}_{R} & -\mathbf{B}_{I}\\
\mathbf{B}_{I} & \mathbf{B}_{R}\\
(\mathbf{c}_{1R}+\mathbf{c}_{2R})^{T} & (\mathbf{c}_{1I}+\mathbf{c}_{2I})^{T}\\
(\mathbf{c}_{2I}-\mathbf{c}_{1I})^{T} & (\mathbf{c}_{1R}-\mathbf{c}_{2R})^{T}
\end{array}
\right)}, \tilde{\mathbf{s}}={\left( \begin{array}{c}
\mathbf{s}_{1}^k\\
\mathbf{s}_{2}^k
\end{array}
\right)},
\end{eqnarray*}
and the subscripts with $R$ and $I$ represent the real and imaginary parts of $\mathbf{B}, \mathbf{c}_1, \mathbf{c}_2$, respectively. The following lemma guarantees non-zero solutions of (\ref{eq:eq14}) with respect to $\tilde{\mathbf{s}}$.
\begin{lemma}\label{le:1}
The $2N\times2N$ matrix $\mathbf{W}$ is rank deficient. Consequently the linear system (\ref{eq:eq14}) has at least one non-zero solution.
\end{lemma}
The detailed proof is shown in Appendix \ref{ap:ap2}. We can obtain a non-zero solution via the following process. Let
\begin{eqnarray} \label{eq:eq3}
\tilde{\mathbf{s}}=\mathbf{V}_{2}^{T}\mathbf{u},
\end{eqnarray}
where $\mathbf{V}_{2}$ comes from the singular value decomposition (SVD) of $\mathbf{W}$:
\begin{eqnarray*}
\mathbf{W}=\mathbf{U}{\left( \begin{array}{cc}
\boldsymbol{\Sigma}_{r}&\mathbf{0}\\
\mathbf{0}&\mathbf{0}
\end{array}
\right)}\mathbf{V}.
\end{eqnarray*}
Here $\mathbf{V}_{2}\in\mathbb{C}^{(2N-r)\times 2N}$ contains the last $2N-r$ rows of the orthogonal matrix $\mathbf{V}$, and $r=\textrm{rank}(\mathbf{W})$. And $\mathbf{s}_k$ is constructed from $\tilde{\mathbf{s}}$. Let
\begin{eqnarray} \label{eq:it}
\mathbf{x}_{k+1}=\mathbf{x}_k+\mathbf{s}_k.
\end{eqnarray}
Insert (\ref{eq:eq3}) into (\ref{eq:eq6.1}) and we have
\begin{eqnarray*}
f(\mathbf{x}_{k+1})=f(\mathbf{x}_{k})+\mathbf{u}^{T}\mathbf{T}\mathbf{u}+2\mathbf{t}^{T}\mathbf{u},
\end{eqnarray*}
where $\mathbf{v}=\mathbf{A}_{1}\mathbf{x}_{k}+\mathbf{a}_{1}$,
\begin{eqnarray*}
\mathbf{T}=\mathbf{V}_{2}\left({\begin{array}{cc}
\mathbf{A}_{1R}&-\mathbf{A}_{1I}\\
\mathbf{A}_{1I}&\mathbf{A}_{1R}
\end{array}}
\right)\mathbf{V}_{2}^{T}
\textrm{ and }
\mathbf{t}=\mathbf{V}_{2}\left( {\begin{array}{c}
\mathbf{v}_{R}\\
\mathbf{v}_{I}
\end{array}}
\right).
\end{eqnarray*}
The following analysis becomes binary. If $\mathbf{T}$ has a non-positive eigenvalue, let $\mathbf{u}$ be the corresponding eigenvector satisfying $\mathbf{t}^{T}\mathbf{u}<0$. Consequently $\mathbf{u}^{T}\mathbf{T}\mathbf{u}+2\mathbf{t}^{T}\mathbf{u}<0$ and $f(\mathbf{x}_{k+1})<f(\mathbf{x}_{k})$. Otherwise $\mathbf{T}$ is positive definite. We have $\mathbf{u}=-\mathbf{T}^{-1}\mathbf{t}$. From the positive definiteness of $\mathbf{T}$, it is easy to see that
\begin{eqnarray*}
f(\mathbf{x}_{k+1})=f(\mathbf{x}_{k})-\mathbf{t}^T\mathbf{T}^{-1}\mathbf{t}<f(\mathbf{x}_{k})
\end{eqnarray*}
holds for this case, too.

With (\ref{eq:eq15}) and the above analysis, the conclusion of the theorem is proved and $\mathbf{x}_{k+1}$ is obtained. $\Box$

\subsubsection{Subproblem for $\mathbf{y}$}
Next, if we fix $\mathbf{x}$, the subproblem for $\mathbf{y}$ is as follows:
\begin{eqnarray}\label{eq:eq24}
&\displaystyle\min_{\mathbf{y}\in\mathbb{C}^{N\times 1}}& \frac{\rho}{2}\mathbf{y}^{H}\mathbf{y}+\mathbf{y}^{H}\big(-\frac{\rho}{2}\mathbf{x}-\frac{\boldsymbol{\mu}}{2}\big)
+\big(-\frac{\rho}{2}\mathbf{x}^{H}-\frac{\boldsymbol{\mu}^{H}}{2}\big)\mathbf{y}\nonumber\\
&\textrm{s.~t.}& |y_{l}|\leq 1,l=1,2,\ldots,N.
\end{eqnarray}
Here all the variables $y_l$ are independent of each other. Then (\ref{eq:eq24}) can be decomposed into $N$ independent subproblems:
\begin{eqnarray*}
&\displaystyle\min_{y_l\in\mathbb{C}}& \frac{\rho}{2}|y_l|^2-\textrm{Re}(\overline{y_l}(\rho x_l+\mu_l))\\
&\textrm{s.~t.}& |y_{l}|\leq 1.
\end{eqnarray*}
Denote $b_l=x_l+\frac{\mu_l}{\rho}$, and the closed form solution of (\ref{eq:eq24}) is:
\begin{eqnarray*}
y_l^*=\left\{
\begin{array}{ll}
b_l,& \textrm{if }|b_l|\leq1,\\
\frac{b_l}{|b_l|},& \textrm{otherwise}.
\end{array}
\right.
\end{eqnarray*}

\subsubsection{Algorithm framework}
Based on the above analysis, the proposed ADMM algorithm for problem (\ref{eq:eq4}) is shown in Alg. 1 \cite{Eckstein}.

\begin{algorithm}\label{alg:1}
\SetKwInOut{Input}{input}
\SetKwInOut{Output}{output}
\caption{ADMM method for weighted sum SINR maximization}

\Input{$k=0$; stopping parameters $\epsilon \ge 0$; the maximum number of iterations $K_{\max}$; initial\\
 \vspace{0.1cm}
 penalty parameter $\rho^{(0)}$ and $\delta>1$; initial variables $\mathbf{x}_0$, $\mathbf{y}_0$ and $\boldsymbol{\mu}^{(0)}$.}
 \vspace{0.1cm}
\Output{$\mathbf{x}_{k}$}
\vspace{0.1cm}
\Repeat{$\max\{\|\mathbf{x}_k-\mathbf{y}_k\|,\|\mathbf{y}_k-\mathbf{y}_{k-1}\|\}\leq \epsilon$ or $k>K_{\max}$}{
\vspace{0.1cm}
1. Solve subproblem (\ref{eq:eq6}) with $\mathbf{y}_k$ to obtain $\mathbf{x}_{k+1}$\;
\vspace{0.1cm}
2. Solve subproblem (\ref{eq:eq24}) with $\mathbf{x}_{k+1}$ to obtain $\mathbf{y}_{k+1}$\;
\vspace{0.1cm}
3. Update the Lagrange multipliers:
\vspace{0.1cm}
$\boldsymbol{\mu}^{(k+1)}:=\boldsymbol{\mu}^{(k)}+\rho^{(k)}(\mathbf{x}_{k+1}-\mathbf{y}_{k+1})$.\\
\vspace{0.1cm}
\If{$\|\mathbf{x}_{k+1}-\mathbf{y}_{k+1}\|_{2}>\frac{1}{4} \|\mathbf{x}_k-\mathbf{y}_k\|_{2}$}{
\vspace{0.1cm}
    Update the penalty factor $\rho^{(k+1)}=\delta\rho^{(k)}$\;
}
\vspace{0.1cm}
$k:=k+1$\;
}
\end{algorithm}

In the algorithm framework, we require that the feasibility violation $\|\mathbf{x}_k-\mathbf{y}_k\|_2$ must have sufficient reduction in each iteration. Thus we can guarantee that the feasibility violation converges to $0$ when we apply Alg. 1 to problem (\ref{eq:eq4}). The maximum iteration number $K_{\max}$ in Alg. 1 is only used for algorithm safeguard. In the numerical tests the algorithm always converges. In the following theorem it is shown that as long as the iteration point in Alg. 1 converges, it converges to the KKT point of problem (\ref{eq:eq4}).
\begin{theorem}\label{theo:4}
Suppose Alg. 1 converges:
$$
\mathbf{x}_k\rightarrow \mathbf{x}^*, \mathbf{y}_k\rightarrow \mathbf{y}^*,
$$
and thus the corresponding Lagrange multiplier and penalty parameter converge, too:
$$
\boldsymbol{\mu}^{(k)}\rightarrow\boldsymbol{\mu}^*, \rho^{(k)}\rightarrow\rho^*,
$$
when $k\rightarrow +\infty$. Suppose $\rho^*$ is sufficiently large, satisfying (\ref{eq:eq76}). Then $\mathbf{x}^*$ is the KKT point of problem (\ref{eq:eq4}).
\end{theorem}
The detailed proof is provided in Appendix \ref{ap:ap3}.


\section{Sum rate maximization model}\label{sec:sr}
In this section, the sum rate is used as the system QoS measurement. By the Shannon capacity, the rate of User $j$ is defined as
$$
\log_2(1+\textrm{SINR}_j), j=1,2.
$$
Replace the objective function of problem (\ref{eq:wSINR}) with the sum rate, then the corresponding sum rate maximization problem is
\begin{eqnarray} \label{eq:SR}
&\displaystyle\max_{\mathbf{w}_{1},\mathbf{w}_{2},\boldsymbol{\Theta}}& R_{\textrm{sum}}=\sum_{j=1}^{2}\textrm{log}_2(1+\textrm{SINR}_j(\mathbf{w}_{1},\mathbf{w}_{2},\boldsymbol{\Theta})) \\
&\textrm{s.~t.}& \|\mathbf{w}_{1}\|^2+\|\mathbf{w}_{2}\|^2\leq P_T,\nonumber\\
&&\boldsymbol{\Theta}=\textrm{Diag}(\theta_{1},\ldots,\theta_{N}),|\theta_{l}|\leq 1,l=1,\ldots,N.\nonumber
\end{eqnarray}

\subsection{New approximation model}
From the discussion in Section \ref{sec:3.2}, the upper bound of SINR is obtained as (\ref{eq:eq8}). And thus we can obtain the upper bound of the sum rate as:
$$
R_{\textrm{sum}}=\textrm{log}_2\left(1+\frac{|\mathbf{h}_1^H\mathbf{w}_1|^2}{|\mathbf{h}_1^H\mathbf{w}_2|^2+\sigma_{1}^2}\right)+
\textrm{log}_2\left(1+\frac{|\mathbf{h}_2^H\mathbf{w}_2|^2}{|\mathbf{h}_2^H\mathbf{w}_1|^2+\sigma_2^2}\right)
\leq\sum_{j=1}^{2}\textrm{log}_2\left(1+\frac{\omega_j^2P_T\|\mathbf{h}_j\|^2}{\sigma_j^2}\right),
$$
where the upper bound is achievable with the two conditions below (\ref{eq:eq8}). Next we maximize the upper bound of $R_{\textrm{sum}}$ instead of itself. With some equivalent transformation, the corresponding optimization problem becomes:
\begin{eqnarray}\label{eq:tx}
&\displaystyle\max_{t\in\mathbb{R},\mathbf{x}\in\mathbb{C}^N}& [\sigma_1^{2}+(1-t)P_T\|\mathbf{d}_{1}
+\mathbf{F}_{1}\mathbf{x}\|^{2}](\sigma_2^{2}+tP_T\|\mathbf{d}_{2}
+\mathbf{F}_{2}\mathbf{x}\|^{2}) \\
&\textrm{s.~t.}& 0\leq t\leq 1,\nonumber\\
&&(\mathbf{d}_{1}+\mathbf{F}_{1}\mathbf{x})^{H}(\mathbf{d}_{2}+\mathbf{F}_{2}\mathbf{x})=0,\nonumber\\
&&|\mathbf{x}_{l}|\leq 1,l=1,\ldots,N.\nonumber
\end{eqnarray}
Here $t=\omega_2^2$ represents the power allocation scheme, and other parameters are defined as in Section \ref{sec:3.2}. Under the same conditions as in Theorem \ref{theo:1}, we can show that problem (\ref{eq:tx}) is feasible. The proof is similar to that of Theorem \ref{theo:1}. And any feasible point of problem (\ref{eq:tx}) is feasible for the sum rate maximization problem (\ref{eq:SR}).

In the objective function of (\ref{eq:tx}), variables $t$ and $\mathbf{x}$ are coupled together. It is nontrivial to solve $t$ and $\mathbf{x}$ jointly. Interestingly, for any feasible $\mathbf{x}$, the problem of $t$ is like:
\begin{eqnarray}
\max_{t\in\mathbb{R}}\hspace{0.2cm}[b_1(1-t)+a_1](b_{2}t+a_2)\hspace{0.2cm}\textrm{s.~t.}\hspace{0.2cm}t\in[0,1],
\end{eqnarray}
with $a_j=\sigma_j^2$, $b_j=P_T\|\mathbf{d}_j+\mathbf{F}_j\mathbf{x}\|^2$, $j=1,2$. It is a one-dimensional quadratic programming problem. We can write down its closed-form solution as:
\begin{eqnarray}\label{eq:topt}
t^*(\mathbf{x})=\left\{
\begin{array}{ll}
\frac{1}{2}+\frac{a_1b_2-a_2b_1}{2b_{1}b_{2}},& \textrm{if }|a_1b_2-a_2b_1|\leq b_{1}b_{2},\\
1,& \textrm{if }a_1b_2-a_2b_1>b_{1}b_{2},\\
0,& \textrm{if }a_1b_2-a_2b_1+b_{1}b_{2}<0.
\end{array}
\right.
\end{eqnarray}
Here $t^*$ is a function of $\mathbf{x}$. When $t^*$ equals $1$ and $0$, the BS concentrates all the power on a single user, which degenerates to the single user case. And the single user results apply \cite{Wu2018}. Otherwise we plug the first case of $t^*$ back into problem (\ref{eq:tx}). with equivalent transformation, a problem of $\mathbf{x}$ is obtained:
\begin{eqnarray}\label{eq:eq13}
&\displaystyle\max_{\mathbf{x}\in\mathbb{C}^N}& P_T^{2}\|\mathbf{d}_{1}+\mathbf{F}_{1}\mathbf{x}\|^{2}\|\mathbf{d}_{2}+\mathbf{F}_{2}\mathbf{x}\|^{2}+
2P_T(\sigma_2^2\|\mathbf{d}_{1}+\mathbf{F}_{1}\mathbf{x}\|^{2}+\sigma_1^2\|\mathbf{d}_{2}+\mathbf{F}_{2}\mathbf{x}\|^{2})\nonumber\\[-0.1cm]
&&+\sigma_1^4\frac{\|\mathbf{d}_{2}
+\mathbf{F}_{2}\mathbf{x}\|^{2}}{\|\mathbf{d}_{1}+\mathbf{F}_{1}\mathbf{x}\|^{2}}
+\sigma_2^4\frac{\|\mathbf{d}_{1}+\mathbf{F}_{1}\mathbf{x}\|^{2}}{\|\mathbf{d}_{2}+\mathbf{F}_{2}\mathbf{x}\|^{2}}\nonumber\\[0.2cm]
&\textrm{s.~t.}& (\mathbf{d}_{1}+\mathbf{F}_{1}\mathbf{x})^{H}(\mathbf{d}_{2}+\mathbf{F}_{2}\mathbf{x})=0,\nonumber\\[-0.1cm]
&&|\mathbf{x}_{l}|\leq 1,l=1,\ldots,N.
\end{eqnarray}
There are sum of ratio terms in the objective function. Thus, problem (\ref{eq:eq13}) is very difficult. If the local noise power at the two users are the same, that is, $\sigma_1^2=\sigma_2^2=\sigma^2$, then we can obtain the constant lower bound of the sum of ratio terms:
$$
\sigma^4\frac{\|\mathbf{d}_{2}
+\mathbf{F}_{2}\mathbf{x}\|^{2}}{\|\mathbf{d}_{1}+\mathbf{F}_{1}\mathbf{x}\|^{2}}
+\sigma^4\frac{\|\mathbf{d}_{1}+\mathbf{F}_{1}\mathbf{x}\|^{2}}{\|\mathbf{d}_{2}+\mathbf{F}_{2}\mathbf{x}\|^{2}}\geq2\sigma^4.
$$
Such hypothesis usually holds for numerical experiments and practical homogenous receivers. Thus the sum of ratio terms are replaced by $2\sigma^4$, and removed from the objective function. The new approximation problem becomes:
\begin{subequations}\label{eq:ap}
\begin{eqnarray}
&\displaystyle\max_{\mathbf{x}\in\mathbb{C}^N}& P_T^{2}\|\mathbf{d}_{1}+\mathbf{F}_{1}\mathbf{x}\|^{2}\|\mathbf{d}_{2}+\mathbf{F}_{2}\mathbf{x}\|^{2}
+2P_T\sigma^2(\|\mathbf{d}_{1}+\mathbf{F}_{1}\mathbf{x}\|^{2}+\|\mathbf{d}_{2}+\mathbf{F}_{2}\mathbf{x}\|^{2})\\
&\textrm{s.~t.}& (\mathbf{d}_{1}+\mathbf{F}_{1}\mathbf{x})^{H}(\mathbf{d}_{2}+\mathbf{F}_{2}\mathbf{x})=0,\label{eq:ap.1}\\
&&|\mathbf{x}_{l}|\leq 1,l=1,\ldots,N.\label{eq:ap.2}
\end{eqnarray}
\end{subequations}
It is simplified significantly compared to the original model (\ref{eq:SR}). However there are still two main difficulties to solve problem (\ref{eq:ap}). First, the objective function is a quartic function of $\mathbf{x}$, which is nonconvex and highly nonlinear. Second, similar to the analysis in Section \ref{sec:ADMM}, there are many constraints which are nonconvex.

\subsection{ADMM algorithm and analysis} \label{sec:ADMM2}
In this subsection, an ADMM based algorithm is proposed for problem (\ref{eq:ap}), which overcomes the two difficulties mentioned above.

First, auxiliary variables $\mathbf{y}$ and $\mathbf{z}$ are introduced. The equivalent problem is formulated:
\begin{eqnarray}\label{eq:eq21}
&\displaystyle\min_{\mathbf{x}, \mathbf{y}, \mathbf{z}}& -P_T^{2}\|\mathbf{d}_{1}+\mathbf{F}_{1}\mathbf{x}\|^{2}\|\mathbf{z}\|^{2}-
2P_T\sigma^2(\|\mathbf{d}_{1}+\mathbf{F}_{1}\mathbf{x}\|^{2}+\|\mathbf{z}\|^{2})\nonumber\\
&\textrm{s.~t.}& (\mathbf{d}_{1}+\mathbf{F}_{1}\mathbf{x})^{H}(\mathbf{d}_{2}+\mathbf{F}_{2}\mathbf{x})=0,\nonumber\\
&&|\mathbf{y}_{l}|\leq 1,l=1,\ldots,N,\nonumber\\
&&\mathbf{y}=\mathbf{x},\nonumber\\
&&\mathbf{z}=\mathbf{d}_{2}+\mathbf{F}_{2}\mathbf{x}.
\end{eqnarray}
Here the auxiliary constraint $\mathbf{y}=\mathbf{x}$ is used, so that different kinds of constraints (\ref{eq:ap.1}) and (\ref{eq:ap.2}) are separated. We use $\mathbf{z}=\mathbf{d}_{2}+\mathbf{F}_{2}\mathbf{x}$ to simplify the complicated objective function.

Then we apply the Augmented Lagrangian penalty function technique to penalize the auxiliary constraints to the objective function. The problem becomes:
\begin{eqnarray} \label{eq:al}
\hspace{-0.3cm}&\displaystyle\min_{\mathbf{x}, \mathbf{y}, \mathbf{z}}& \hspace{-0.1cm} -P_T^{2}\|\mathbf{d}_{1}+\mathbf{F}_{1}\mathbf{x}\|^{2}\|\mathbf{z}\|^{2}
-2P_T\sigma^2(\|\mathbf{d}_{1}+\mathbf{F}_{1}\mathbf{x}\|^{2}+\|\mathbf{z}\|^{2})
+\textrm{Re}[\boldsymbol{\mu}_{1}^{H}(\mathbf{x}-\mathbf{y})]\nonumber\\
\hspace{-0.3cm}&&\hspace{-0.1cm}+\frac{\rho_{1}}{2}\|\mathbf{x}-\mathbf{y}\|^{2}
+\textrm{Re}[\boldsymbol{\mu}_{2}^{H}(\mathbf{F}_{2}\mathbf{x}+\mathbf{d}_{2}-\mathbf{z})]
+\frac{\rho_{2}}{2}\|\mathbf{F}_{2}\mathbf{x}+\mathbf{d}_{2}-\mathbf{z}\|^{2}\nonumber\\
\hspace{-0.3cm}&\textrm{s.~t.}&\hspace{-0.1cm} (\mathbf{d}_{1}+\mathbf{F}_{1}\mathbf{x})^{H}(\mathbf{d}_{2}+\mathbf{F}_{2}\mathbf{x})=0,\nonumber\\
\hspace{-0.3cm}&&\hspace{-0.1cm}|\mathbf{y}_{l}|\leq 1,l=1,\ldots,N,
\end{eqnarray}
where $\boldsymbol{\mu}_j$, $j=1,2$ are the corresponding Lagrange multipliers, and $\rho_j$, $j=1,2$ are the penalty parameters.

Next, we solve $\mathbf{x}$, $\mathbf{y}$ and $\mathbf{z}$ in problem (\ref{eq:al}) alternatively. If we fix $\mathbf{y}$ and $\mathbf{z}$, the subproblem for $\mathbf{x}$ has the same expression as (\ref{eq:eq6}), but with different parameters. Here
\begin{eqnarray} \label{eq:a}
\mathbf{A}_1=-P_T(P_T||\mathbf{z}||^{2}+2\sigma^2)\mathbf{F}_{1}^{H}\mathbf{F}_{1}+\frac{\rho_{1}}{2}\mathbf{I}_{N}
+\frac{\rho_{2}}{2}\mathbf{F}_{2}^{H}\mathbf{F}_{2},\\
 \mathbf{a}_1=-P_T(P_T||\mathbf{z}||^{2}+2\sigma^2)\mathbf{F}_{1}^{H}\mathbf{d}_{1}+\frac{1}{2}\boldsymbol{\mu}_{1}
+\frac{1}{2}\mathbf{F}_{2}^{H}\boldsymbol{\mu}_{2}-\frac{\rho_{1}}{2}\mathbf{y}+\frac{\rho_{2}}{2}\mathbf{F}_{2}^{H}(\mathbf{d}_{2}-\mathbf{z}),
\end{eqnarray}
and $\mathbf{A}_2$, $\mathbf{a}_2$, $\mathbf{a}_3$ and $a_4$ are defined below (\ref{eq:eq6}).
Obviously, if $\rho_1$ is sufficiently large, that is,
\begin{eqnarray}\label{eq:p1}
\rho_1>2\lambda_{\textrm{max}}(P_T(P_T\|\mathbf{z}\|^{2}+2\sigma^2)\mathbf{F}_{1}^{H}\mathbf{F}_1-\rho_2\mathbf{F}_2^H\mathbf{F}_2),
\end{eqnarray}
then subproblem (\ref{eq:eq6}) is guaranteed to be bounded below. The optimal solution can be achieved through the SDR method. Otherwise similar to the analysis in Section \ref{sec:x}, we solve a linear equation to obtain a new feasible iteration point with sufficient reduction of the objective function. The analysis and theorem in Section \ref{sec:x} works for this subproblem, too.

The subproblem for $\mathbf{y}$ is in the following form:
\begin{eqnarray}\label{eq:y}
&\displaystyle\min_{\mathbf{y}\in\mathbb{C}^{N\times 1}}& \frac{\rho_1}{2}\mathbf{y}^{H}\mathbf{y}-\textrm{Re}\big(\mathbf{y}^{H}(\rho_1\mathbf{x}+\boldsymbol{\mu}_1)\big)\nonumber\\
&\textrm{s.~t.}& |y_{l}|\leq 1,l=1,2,\ldots,N.
\end{eqnarray}
Each element $y_l$ is independent of each other in (\ref{eq:y}). Let $\mathbf{b}=\mathbf{x}+\frac{\boldsymbol{\mu}_1}{\rho_1}$. Its closed form solution is
\begin{eqnarray}
y_l^*=\left\{
\begin{array}{ll}
b_l,& \textrm{if }|b_l|\leq1,\\
\frac{b_l}{|b_l|},& \textrm{otherwise},
\end{array}
\right.
\hspace{0.2cm}l=1,2,\ldots,N.
\end{eqnarray}

Fixing $\mathbf{x}$ and $\mathbf{y}$, we have the subproblem for $\mathbf{z}$ as:
\begin{eqnarray}\label{eq:z}
&\displaystyle\min_{\mathbf{z}\in\mathbb{C}^{M}}& h(\mathbf{z})=\frac{c}{2}\|\mathbf{z}\|^2+\textrm{Re}(\mathbf{z}^{H}\mathbf{p}),
\end{eqnarray}
where $c=\rho_{2}-2P_T(P_T\|\mathbf{d}_{1}+\mathbf{F}_{1}\mathbf{x}\|^2+2\sigma^2)$ and $\mathbf{p}=-\boldsymbol{\mu}_2-\rho_2(\mathbf{F}_{2}\mathbf{x}+\mathbf{d}_{2})$. If
\begin{eqnarray}\label{eq:p2}
\rho_2>2P_T(P_T\|\mathbf{d}_{1}+\mathbf{F}_{1}\mathbf{x}\|^2+\sigma^2),
\end{eqnarray}
that is $c>0$, the unconstrained subproblem (\ref{eq:z}) is convex and bounded. Then its closed form solution is $\mathbf{z}^*=-\frac{\mathbf{p}}{c}$. Otherwise $c\leq0$, (\ref{eq:z}) is unbounded. Then we just go along the gradient step with unit stepsize, to obtain a new iteration point with sufficient reduction of function value. From the analysis, the iteration formula is as follows:
\begin{eqnarray} \label{eq:zs}
\mathbf{z}_{k+1}=\left\{
\begin{array}{ll}
-\frac{\mathbf{p}}{c},& \textrm{if }c>0,\\
\mathbf{z}_k-\mathbf{p},& \textrm{if }c=0,\\
(1-c)\mathbf{z}_k-\mathbf{p},& \textrm{if }c<0.
\end{array}
\right.
\end{eqnarray}

With the above analysis, the ADMM based algorithm framework is shown as Alg. 2.
\begin{algorithm}\label{alg:2}
\SetKwInOut{Input}{input}
\SetKwInOut{Output}{output}
\caption{ADMM method for sum rate maximization}

\Input{$k=0$; stopping parameters $\epsilon \ge 0$; the maximum number of iterations $K_{\max}$; initial\\
 \vspace{0.1cm}
 penalty parameter $\rho_j^{(0)}$, $j=1,2$ and $\delta>1$; initial variables $\mathbf{x}_0$, $\mathbf{y}_0$, $\mathbf{z}_0$ and $\boldsymbol{\mu}_j^{(0)}$,\\
 \vspace{0.1cm}
  $j=1,2$.}
  \vspace{0.1cm}
\Output{$\mathbf{x}_{k}$}
\vspace{0.1cm}
\Repeat{$\max\{\|\mathbf{x}_k-\mathbf{y}_k\|,\|\mathbf{F}_2\mathbf{x}_k+\mathbf{d}_2-\mathbf{z}_k\|,\|\mathbf{y}_k-\mathbf{y}_{k-1}\|,
    \|\mathbf{z}_k-\mathbf{z}_{k-1}\|\}\leq \epsilon$ or $k>K_{\max}$}{
    \vspace{0.1cm}
1. Solve Subproblem (\ref{eq:eq6}) with $\mathbf{y}_k$ and $\mathbf{z}_k$ to obtain $\mathbf{x}_{k+1}$\;
\vspace{0.1cm}
2. Solve Subproblem (\ref{eq:y}) with $\mathbf{x}_{k+1}$ and $\mathbf{z}_k$ to obtain $\mathbf{y}_{k+1}$\;
\vspace{0.1cm}
3. Solve Subproblem (\ref{eq:z}) with $\mathbf{x}_{k+1}$ and $\mathbf{y}_{k+1}$ to obtain $\mathbf{z}_{k+1}$\;
\vspace{0.1cm}
4. Update the Lagrange multipliers:
\vspace{0.1cm}
$\boldsymbol{\mu}_1^{(k+1)}=\boldsymbol{\mu}_1^{(k)}+\rho_1^{(k)}(\mathbf{x}_{k+1}-\mathbf{y}_{k+1})$, \vspace{0.1cm}
$\boldsymbol{\mu}_2^{(k+1)}=\boldsymbol{\mu}_2^{(k)}+\rho_2^{(k)}(\mathbf{F}_{2}\mathbf{x}_{k+1}+\mathbf{d}_{2}-\mathbf{z}_{k+1})$\;
\vspace{0.1cm}
\If{$\|\mathbf{x}_{k+1}-\mathbf{y}_{k+1}\|>\frac{1}{4} \|\mathbf{x}_{k}-\mathbf{y}_k\|$}{
\vspace{0.1cm}
    Update the penalty factor $\rho_1^{(k+1)}=\delta_1\rho_1^{(k)}$\;
}
\vspace{0.1cm}
\If{$\|\mathbf{F}_2\mathbf{x}_{k+1}+\mathbf{d}_2-\mathbf{z}_{k+1}\|>\|\mathbf{F}_2\mathbf{x}_k+\mathbf{d}_2-\mathbf{z}_k\|$}{
\vspace{0.1cm}
Update the penalty factor $\rho_2^{(k+1)}=\delta_2\rho_2^{(k)}$\;
}
\vspace{0.1cm}
$k:=k+1$\;
}
\end{algorithm}

The theoretical analysis of Alg. 2 is similar to that of Alg. 1. The following theorem holds.
\begin{theorem}\label{theo:2}
Suppose Alg. 2 converges:
$$
\mathbf{x}_k\rightarrow \mathbf{x}^*, \mathbf{y}_k\rightarrow \mathbf{y}^* \textrm{ and } \mathbf{z}_k\rightarrow \mathbf{z}^*,
$$
and thus the corresponding Lagrange multipliers and penalty parameters converge too:
$$
\boldsymbol{\mu}_j^{(k)}\rightarrow\boldsymbol{\mu}_j^*, \rho_j^{(k)}\rightarrow\rho_j^*, j=1,2,
$$
when $k\rightarrow +\infty$. Suppose $\rho_1^*$ and $\rho_2^*$ are sufficiently large, satisfying (\ref{eq:p1}) and (\ref{eq:p2}). Then $\mathbf{x}^*$ is the KKT point of problem (\ref{eq:ap}).
\end{theorem}
The proof is a straightforward extension of that of Theorem \ref{theo:4}. Thus the detailed proof is omitted in this part.

\subsection{Discussions}
In this paper, we mainly consider the communication model with two users. The main interference cancellation technique can be extended for $J(>2)$ number of users. First, the equivalent channels from the BS to different users should be orthogonal:
$$
\mathbf{h}_j^H\mathbf{h}_m=0, j\neq m, \forall j, m=1,\ldots, J,
$$
where $\mathbf{h}_j=\mathbf{d}_j+\mathbf{F}_j\mathbf{x}$ is the equivalent channel from BS to user j. This requires $\frac{J(J-1)}{2}\leq N$, so that the above nonlinear system is feasible. Second, all the precoding vectors are applying the MRT precoders:
$$
\mathbf{w}_j=\omega_j\sqrt{P_T}\frac{\mathbf{h}_j}{\|\mathbf{h}_j\|_2}, j=1,\ldots, J,
$$
where $\sum_{j=1}^J\omega_j^2\leq1$. Then the SINR of each user can achieve its upper bound
$$
\textrm{SINR}_j\leq \frac{\omega_j\sqrt{P_T}\|\mathbf{h}_j\|_2^2}{\sigma_j^2}, j=1,\ldots, J.
$$
For the weighted SINR maximization model, the approximated problem is still a QCQP problem. Optimization techniques for QCQP problems can be applied. For the sum rate maximization model, the objective function becomes highly nonlinear with $J$ greater than $2$, which requires more specific techniques for simplification.

The weighted SINR maximization model (\ref{eq:eq4}) adopts equal transmit power distribution to avoid single-user case degeneration. In fact, multi-user transmission can also be achieved by extra SINR requirement for each user. That is, if we add SINR constraints for the two users to problem (\ref{eq:eq16}), it will be well-defined. Furthermore, sum rate maximization model with both transmit power constraint and SINR requirements are reasonable, too. These problems are beyond the scope of this paper, and will be left for future work.

\section{Simulation results}
In this section, we shall verify the effectiveness of the proposed models as well as algorithms via numerical experiments. Each presented data is the average result of $1000$ random realizations if there is no further explanation, and all the generated channels are Rician fading channels as described below.

\subsection{Weighted sum SINR maximization}\label{sec:simu1}
The weighted sum SINR maximization model is considered and tested in this part. The experiment setting is similar to \cite{Wu}. The antenna number of the BS is $M=40$ and the number of RIS elements is $N=80$. The transmit power of the BS is $P_T=2$. The local noise power at each user is set as $\sigma^2=\sigma_1^2=\sigma_2^2=-80$dBm. We generate the channel from the BS to the RIS following the rule
$$
\mathbf{F}=\sqrt{\frac{\beta_{BR}}{1+\beta_{BR}}}\mathbf{F}^{LoS}+\sqrt{\frac{1}{1+\beta_{BR}}}\mathbf{F}^{NLoS},
$$
where $\mathbf{F}^{LoS}$ is a matrix with all elements as $1$; all elements in $\mathbf{F}^{NLoS}$ obey complex Gaussian distribution $\mathcal{CN}(0,1)$; the Racian factor $\beta_{BR}=3$dB. The channels from the BS and the RIS to users are generated similarly, with Racian factors $\beta_{BU}=\beta_{RU}=3$dB. The parameters in Alg. 1 are set as $\epsilon=10^{-5}$, $\rho_{0}=10^{-4}$, $\delta=1.01$ and $K_{\max}=100$. The initial point $\mathbf{x}_0$ satisfies constraint (\ref{eq:eq4.1}), and $\mathbf{y}_0$ and $\boldsymbol{\mu}^{(0)}$ are randomly selected.

\begin{figure}[t]
\begin{minipage}[b]{1.0\linewidth}
\centering
\centerline{\includegraphics[width=10cm]{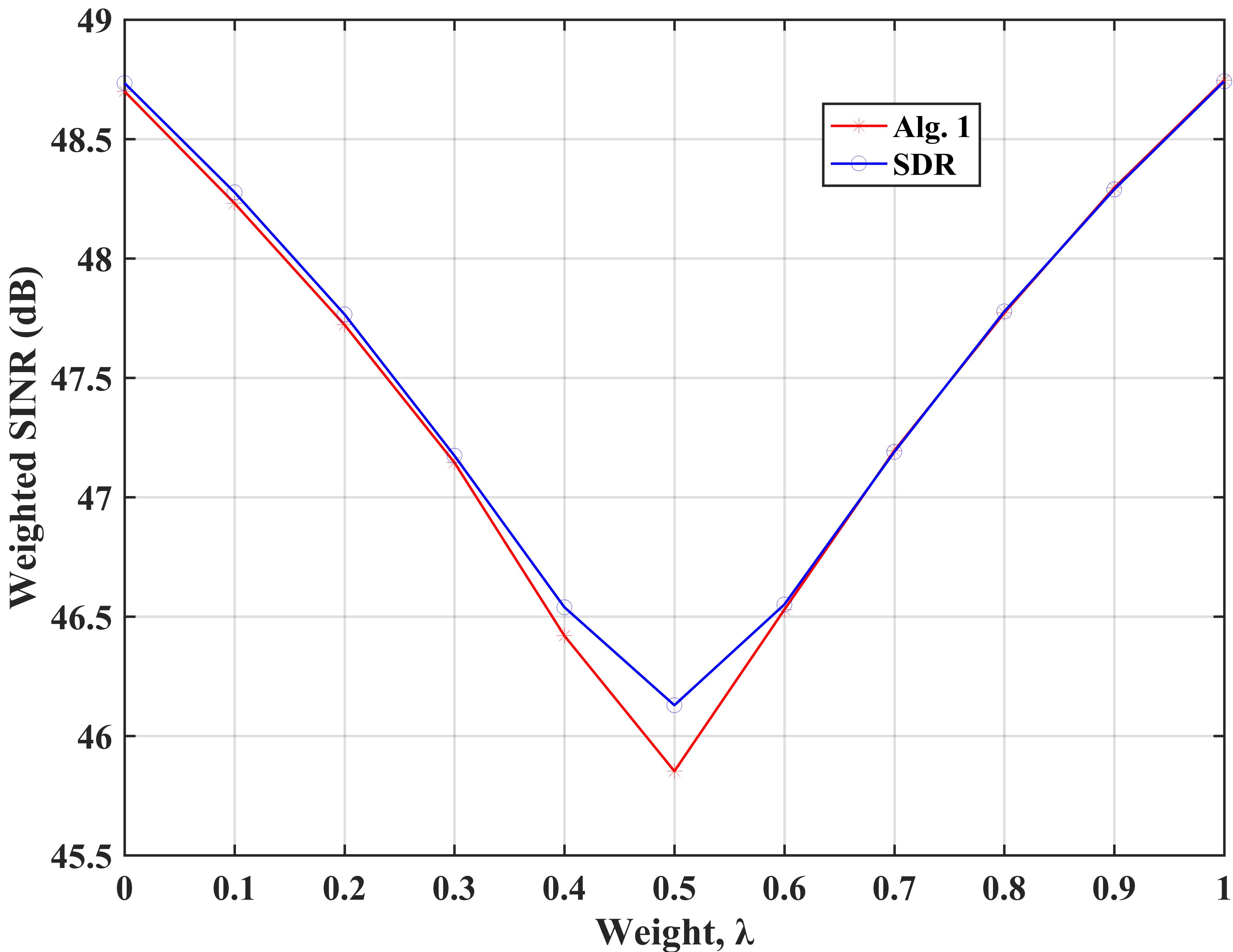}}
\caption{Comparison between Alg. 1 and SDR with\\ $M=40$, $N=80$, $P_T=2$ and $\sigma^2=-80$dBm}
\label{fig2}
\end{minipage}
\end{figure}
First, we compare the proposed Alg. 1 with the classic SDR method for solving QCQP problem (\ref{eq:eq4}). The achieved weighted sum SINR value with respect to weight parameter $\lambda$ are listed in Fig. 2. The two curves almost coincide on the two sides, while the result of Alg. 1 is lower than that of SDR for about 1\% at $\lambda=0.5$. In fact, such comparison is not fair to Alg. 1. As shown in Table 1, ADMM solves all the $1000$ test examples for each value of $\lambda$ while SDR cannot. Especially when $\lambda=0.5$, only $40$ out of $1000$ realizations are solved by SDR, and it fails to find a feasible solution for other realizations. The result of SDR with respect to $\lambda=0.5$ is only the average result of the $40$ examples. Compared to SDR, Alg. 1 is much more robust with very little performance loss. Thus, it is reasonable to use the proposed Alg. 1 to solve problem (\ref{eq:eq4}).

\begin{table}[htb]
  \centering
  \begin{tabular}{cccccc}
     \hline
     $\lambda$&0 - 0.3&0.4&0.5&0.6&0.7 - 1 \\
     \hline
     ADMM&1000&1000&1000&1000&1000\\
     \hline
     SDR&1000&868&40&873&1000\\
     \hline
  \end{tabular}
  \caption{Number of solved problems in $1000$ realizations}
  \label{table1}
\end{table}

Next, four different models are compared with the proposed model (\ref{eq:eq4}) and Alg. 1. The four referred models are listed below.\\
1. \emph{No RIS exists in the network}\\
In this case, the zero-forcing precoding technique is applied, with
$(\mathbf{w}_1,\mathbf{w}_2)=(\mathbf{d}_1,\mathbf{d}_2)^{+}.$

\vspace{0.2cm}
\noindent2. \emph{RIS with random phase shifts}\\
Each phase shift $\tau_l$ of RIS element $\theta_l=e^{i\tau_l}$ is uniformly generated from $[0,2\pi)$. Then the following problem requires solving.
\begin{eqnarray} \label{eq:eq12}
&\displaystyle\max_{\mathbf{w}_{1},\mathbf{w}_{2}}& \lambda\textrm{SINR}_{1}(\mathbf{w}_{1},\mathbf{w}_{2})+(1-\lambda)
\textrm{SINR}_{2}(\mathbf{w}_{1},\mathbf{w}_{2})\nonumber\\
&\textrm{s.~t.}& \|\mathbf{w}_{1}\|\leq 1, \|\mathbf{w}_{2}\|\leq 1,
\end{eqnarray}
where $\textrm{SINR}_{1}=\frac{|\mathbf{h}_1^H\mathbf{w}_1|^2}{|\mathbf{h}_1^H\mathbf{w}_2|^2+\sigma_1^2}$ and $\textrm{SINR}_2=\frac{|\mathbf{h}_2^H\mathbf{w}_2|^2}{|\mathbf{h}_2^H\mathbf{w}_1|^2+\sigma_2^2}$. Two auxiliary variables $t_1$ and $t_2$ are introduced, and problem (\ref{eq:eq12}) is equivalent to
\begin{eqnarray*}
&\displaystyle\max_{\mathbf{w}_{1},\mathbf{w}_{2},t_1,t_2}& \lambda t_1+(1-\lambda)t_2\\
&\textrm{s.~t.}& \textrm{SINR}_{j}(\mathbf{w}_{1},\mathbf{w}_{2})\geq t_j, j=1,2,\\
&&\|\mathbf{w}_{1}\|\leq 1, \|\mathbf{w}_{2}\|\leq 1.
\end{eqnarray*}
Then we solve problem (\ref{eq:eq12}) optimally combining bi-section and SDR techniques.

\vspace{0.2cm}
\noindent3. \emph{RIS with special phase shifts}\\
The discrete Fourier transformation based phase shifts are applied, where
$\theta_l=e^{-\frac{(l-1)^2}{M}i}$, $l=$\\
$1,\ldots,N$.
Similar to the second model, problem (\ref{eq:eq12}) is solved after RIS matrix is determined.

\vspace{0.2cm}
\noindent4. \emph{Optimized RIS matrix}\\
The weighted sum rate maximization model proposed in \cite{Guo} is applied, too.

\begin{figure}[t]
\begin{minipage}[b]{1.0\linewidth}
\centering
\centerline{\includegraphics[width=10cm]{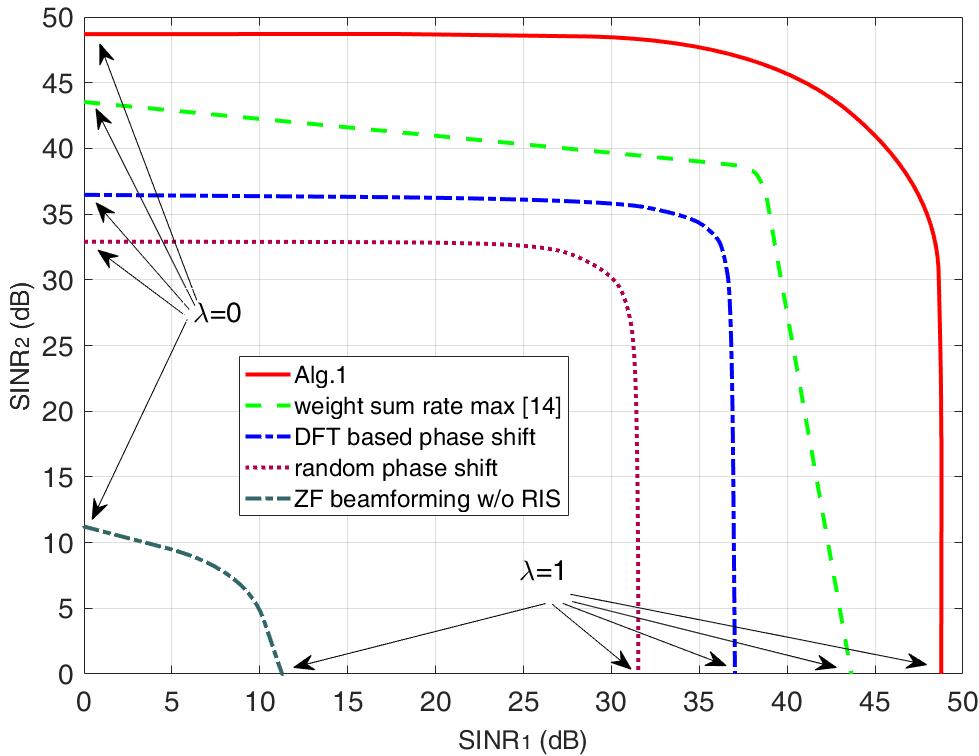}}
\caption{Achievable SINR region with $M=40$, $N=80$, $P_T=2$ and $\sigma^2=-80$dBm}
\label{fig3}
\end{minipage}
\end{figure}

Same randomly generated initial points are used for all five models. And their achieved SINR regions are depicted in Fig. \ref{fig3}.
The sharp contrast indicates that the two models optimizing RIS clearly achieve much higher SINR than the models without RIS and with fixed RIS coefficients. Furthermore, our proposed model outperforms \cite{Guo} in terms of the SINR region.

\subsection{Sum rate maximization}
In this subsection, the proposed sum rate maximization model as well as Alg. 2 are analyzed. The same channel settings as \cite{Guo} are used for fair comparison. The BS has $M=4$ antennas and there are $N=10$ elements in the RIS. The BS and the two users lie on the same line, where the distances between them are $190$m and $210$m respectively. The channel coefficients are generated in the following way. Take the BS-RIS channel $\mathbf{F}$ as an instance:
$$
\mathbf{F}=\kappa_F\mathbf{F}^{LoS}+\mathbf{F}^{NLoS}.
$$
Here $\mathbf{F}^{LoS}$ and $\mathbf{F}^{NLoS}$ are generated in the same way as that in Section \ref{sec:simu1}. $\kappa_F=C_F{d_F}^{-\rho_F}$, where $C_F=-20$dB, $d_F$ denotes the distance between the BS and the RIS, and $\rho_F=2$ is the path loss exponent. Other channel coefficients are generated similarly. The parameters of the direct links are $C_d=-30$dB and $\rho_d=3.5$; those of the RIS-user links are $C_g=-20$dB and $\rho_g=2$. The BS transmit power is set as unit power, and the local noise power at each user is $\sigma^2=\sigma_1^2=\sigma_2^2=-117$dBm. In the proposed Alg. 2, $\epsilon=10^{-4}$, $\rho_1^{(0)}=\rho_2^{(0)}=0.01$, $K_{\max}=100$ and $\delta=5$. The initial point $\mathbf{x}_0$ satisfies constraint (\ref{eq:ap.1}), while $\mathbf{y}_0$, $\mathbf{z}_0$ and $\boldsymbol{\mu}_j^{(0)}, j=1,2$ are randomly selected.

\begin{figure}[t]
\begin{minipage}[b]{1.0\linewidth}
\centering
\centerline{\includegraphics[width=10cm]{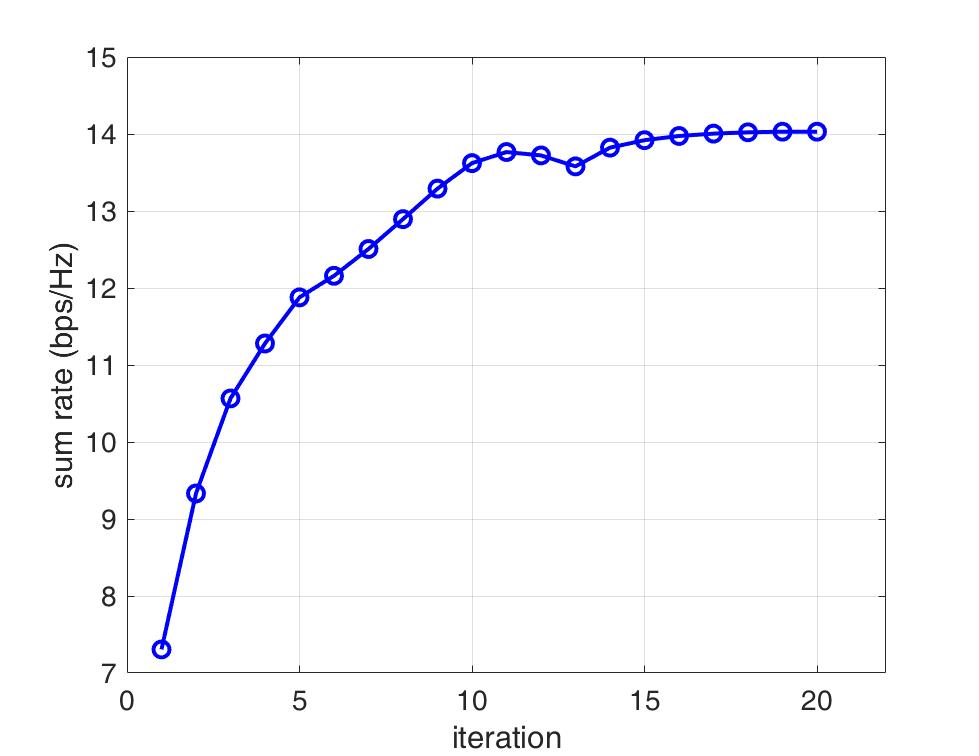}}
\caption{Iteration curve of Alg. 2 with $M=4$, $N=10$, $P_T=1$ and $\sigma^2=-117$dBm}
\label{fig2}
\end{minipage}
\end{figure}
First the iteration curve of the proposed Alg. 2 is presented. Fig. 4 shows its achieved sum rate with respect to iteration number. Basically the sum rate increases and converges in dozens of iterations. The iteration points are usually infeasible for the original problem and thus the reduction of constraint violation results in the nonmonotonic behavior of the achieved sum rate.

\begin{figure}[t]
\begin{minipage}[b]{1.0\linewidth}
\centering
\centerline{\includegraphics[width=10cm]{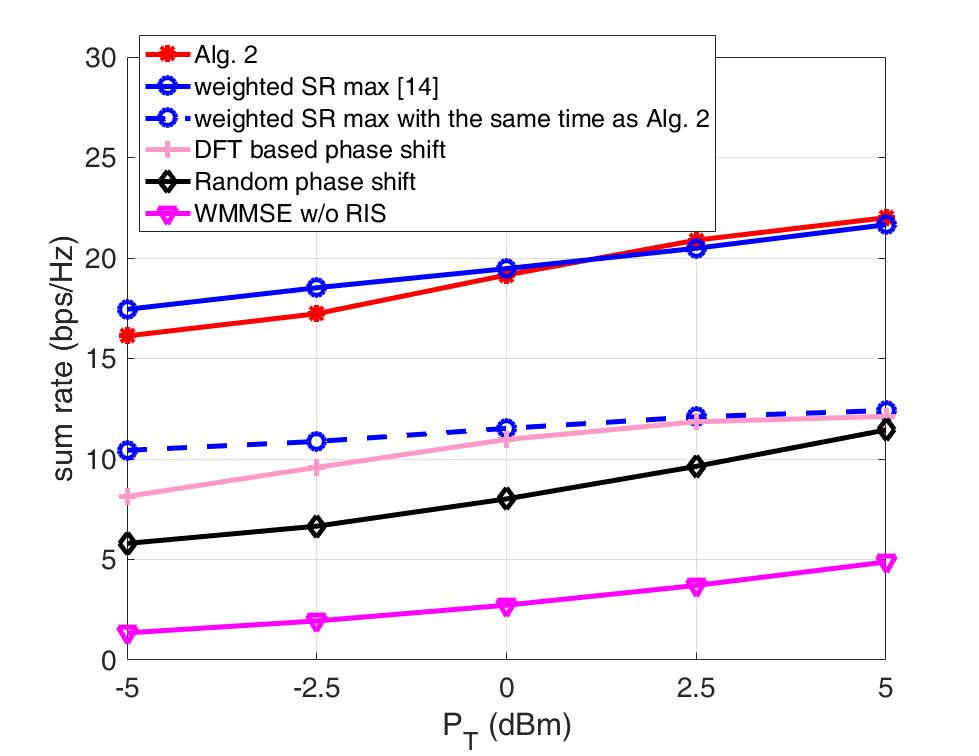}}
\caption{Achievable sum rate with $M=4$, $N=10$, $P_T=1$ and $\sigma^2=-117$dBm}
\label{fig2}
\end{minipage}
\end{figure}
Next we compare the proposed sum rate maximization model as well as Alg. 2 with four different models, which are listed in Section \ref{sec:simu1}. In the first compared model without RIS, we use the classic weighted minimum mean square error (WMMSE) technique \cite{Shi} to solve the sum rate maximization problem. For the second and third compared model, they both fix the RIS parameters and then consider the following problem:
\begin{eqnarray*}
&\displaystyle\max_{\mathbf{w}_{1},\mathbf{w}_{2}}& \sum_{j=1}^2\log_2(1+\textrm{SINR}_j(\mathbf{w}_{1},\mathbf{w}_{2}))\nonumber\\
&\textrm{s.~t.}& \|\mathbf{w}_{1}\|^2+\|\mathbf{w}_{2}\|^2\leq P_T.
\end{eqnarray*}
Similar to problem (\ref{eq:eq12}), we also apply the bi-section technique to solve it optimally. In the fourth compared model we set its weight parameters for different users' rate to be equal, so that to maximize the system sum rate directly. All the compared models use the same random initialization. The achieved sum rate of all models with respect to different transmit power are depicted in Fig. 5 as solid lines. It is observed that the models with fixed RIS parameters have higher sum rate than that without RIS, and they are all surpassed by the models with optimized RIS parameters. The weighted sum rate maximization (WSR) model and the proposed model perform similarly. But they have distinct computational cost. The former applies the BCD technique to alternatively solve the precoders $\mathbf{w}_1$, $\mathbf{w}_2$ and the RIS matrix $\boldsymbol{\Theta}$, and uses ADMM method to solve the subproblem for $\boldsymbol{\Theta}$ in each outer loop. Comparably, the proposed new approximation model (\ref{eq:ap}) only uses ADMM once and solve the whole problem. Thus it has much less computational cost than the WSR model. To have fair comparison, we also plot the achieved sum rate of the WSR model with the same computational time as Alg. 2, which is shown as the dash line in Fig. 5. Clearly our proposed model as well as algorithm greatly outperform the WSR model with the same computational cost. This verifies the efficiency of the proposed model compared to the state of the arts.

\section{Conclusion}
This paper considers the RIS-assisted two-user MISO broacast channel. Two models, the weighted sum SINR maximization and sum rate maximization, are set up with different users' QoS metrics. First, we show the achieved upper bound of the SINR of each user by applying the MRT precoding and ZF techniques. Then the two models are simplified accordingly. The weighted sum SINR maximization model is concluded as a QCQP problem, and the sum rate maximization is further approximated as a bi-quadratic optimization problem with quadratic constraints. Two ADMM based algorithms are proposed for the approximated problems, where each subproblem has either closed form solution or solution with sufficient reduction. The convergence property of the proposed algorithms are analyzed. Simulation results show that the proposed models as well as algorithms are effective, and well balance between system performance and computational cost.

\section{Appendix}  \label{sec:sec5}

\subsection{Proof of Theorem \ref{theo:1}} \label{ap:ap1}
First suppose that the conditions hold when $j=2$. That is, $\|\mathbf{F}_{2}^{+}\mathbf{d}_{2}\|_{\infty}\leq 1$ and $\textrm{rank}(\mathbf{F}_{2})=M$. Let
\begin{eqnarray*}
\mathbf{A}=\left( {\begin{array}{*{20}{c}}
\mathbf{F}_{1}^{H}\mathbf{F}_{2}\\
\mathbf{d}_{1}^{H}\mathbf{F}_{2}
\end{array}} \right)=\left( {\begin{array}{*{20}{c}}
\mathbf{F}_{1}^{H}\\
\mathbf{d}_{1}^{H}
\end{array}} \right)\mathbf{F}_{2},
\end{eqnarray*}
where $\mathbf{F}_j\in \mathbb{C}^{M\times N}$, $\mathbf{d}_j\in \mathbb{C}^{M\times 1}$, $j=1,2$. Thus the constructed matrix $\mathbf{A}$ is an $(N+1)\times N$ matrix. Since the rank of matrix multiplication is no bigger than that of each matrix, we have
$$
\textrm{rank}(\mathbf{A})\leq\min\{
\textrm{rank}\left( {\begin{array}{*{20}{c}}
\mathbf{F}_{1}^{H}\\
\mathbf{d}_{1}^{H}
\end{array}} \right)
,\textrm{rank}(\mathbf{F}_{2})\}\leq M.
$$
As $N\gg M$, the linear system $\mathbf{Az}=\mathbf{0}$ has non-zero solutions generally. Let $\mathbf{y}$ be a nonzero solution of $\mathbf{Az=0}$. Thus, $a\frac{\mathbf{y}}{\|\mathbf{y}\|_{2}}$ is also its solution, for any $a\in \mathbb{R}$:
\begin{subequations}\label{eq:eq30}
\begin{numcases}{}
\mathbf{F}_{1}^{H}\mathbf{F}_{2}\frac{a\mathbf{y}}{\|\mathbf{y}\|_{2}}=\mathbf{0},\\[0.1cm]
\mathbf{d}_{1}^{H}\mathbf{F}_{2}\frac{a\mathbf{y}}{\|\mathbf{y}\|_{2}}=0.
\end{numcases}
\end{subequations}
Let
\begin{eqnarray} \label{eq:eq40}
\mathbf{w}=a\frac{\mathbf{y}}{\|\mathbf{y}\|_{2}}-\mathbf{F}_{2}^{+}\mathbf{d}_{2}, a\in[0,1-\|\mathbf{F}_{2}^{+}\mathbf{d}_{2}\|_{\infty}].
\end{eqnarray}
Since $\textrm{rank}(\mathbf{F}_{2})=M$, $\mathbf{F}_{2}\mathbf{F}_{2}^{+}=\mathbf{I}_{M}$ holds. Take (\ref{eq:eq40}) into (\ref{eq:eq30}), it becomes:
$$
\left\{ \begin{array}{l}
\mathbf{F}_{1}^{H}\mathbf{F}_{2}\mathbf{w}+\mathbf{F}_{1}^{H}\mathbf{d}_{2}=\mathbf{0},\\
\mathbf{d}_{1}^{H}\mathbf{F}_{2}\mathbf{w}+\mathbf{d}_{1}^{H}\mathbf{d}_{2}=0,
\end{array} \right.
$$
which can be rewritten as:
\begin{eqnarray*}
\left( {\begin{array}{*{20}{c}}
\mathbf{F}_{1}^{H}\mathbf{F}_{2}&\mathbf{F}_{1}^{H}\mathbf{d}_{2}\\
\mathbf{d}_{1}^{H}\mathbf{F}_{2}&\mathbf{d}_{1}^{H}\mathbf{d}_{2}
\end{array}} \right) \left( {\begin{array}{*{20}{c}}
\mathbf{w}\\
1
\end{array}} \right)=\mathbf{0}.
\end{eqnarray*}
Obviously, it holds that
\begin{eqnarray*}
\mathbf{w}^{H}\mathbf{F}_{1}^{H}\mathbf{F}_{2}\mathbf{w}+\mathbf{w}^{H}\mathbf{F}_{1}^{H}\mathbf{d}_{2}
+\mathbf{d}_{1}^{H}\mathbf{F}_{2}\mathbf{w}+\mathbf{d}_{1}^{H}\mathbf{d}_{2}=0.
\end{eqnarray*}
Thus, $\mathbf{w}$ satisfies the constraint (\ref{eq:eq4.1}).

Next, according to (\ref{eq:eq40}), we have
\begin{eqnarray*}
\|\mathbf{w}\|_{\infty} \leq \Big{\|}\frac{a\mathbf{y}}{\|\mathbf{y}\|_{2}}\Big{\|}_{\infty}+\|\mathbf{F}_{2}^{+}\mathbf{d}_{2}\|_{\infty} \leq a+\|\mathbf{F}_{2}^{+}\mathbf{d}_{2}\|_{\infty} \leq 1.
\end{eqnarray*}
Thus, $\mathbf{w}$ also satisfies (\ref{eq:eq4.2}). $\mathbf{w}$ is feasible for problem (\ref{eq:eq4}).

Since the constraint (\ref{eq:eq4.1}) can be rewritten as
$$
(\mathbf{d}_{2}+\mathbf{F}_{2}\mathbf{x})^{H}(\mathbf{d}_{1}+\mathbf{F}_{1}\mathbf{x})=0,
$$
the theorem holds with $\|\mathbf{F}_{1}^{+}\mathbf{d}_{1}\|_{\infty}\leq 1$ and $\textrm{rank}(\mathbf{F}_{1})=M$, too.
$\Box$

\subsection{Proof of Lemma \ref{le:1}} \label{ap:ap2}
According to (\ref{eq:eq7.1}), we have the real and imaginary parts of $\mathbf{B}$ as
\begin{eqnarray*}
\mathbf{B}_{R}= \mathbf{F}_{1R}^{T}\mathbf{F}_{2R}+\mathbf{F}_{1I}^{T}\mathbf{F}_{2I}, \mathbf{B}_{I}=\mathbf{F}_{1R}^{T}\mathbf{F}_{2I}-\mathbf{F}_{1I}^{T}\mathbf{F}_{2R}.
\end{eqnarray*}
\noindent Then it is easy to obtain
\begin{eqnarray*}
(\mathbf{B}_R,-\mathbf{B}_I)=(\mathbf{F}_{1R}^T,\mathbf{F}_{1I}^T)
\left( {\begin{array}{cc}
        {{\mathbf{F}_{2R},}}&{ - {\mathbf{F}_{2I}}}\\
        {{\mathbf{F}_{2I},}}&{{\mathbf{F}_{2R}}}
        \end{array}} \right)=\mathbf{AP},
\end{eqnarray*}

\vspace{-0.2cm}
\begin{eqnarray*}
(\mathbf{B}_I,\mathbf{B}_R)=(\mathbf{F}_{1R}^T,\mathbf{F}_{1I}^T)
\left( {\begin{array}{cc}
        {{\mathbf{F}_{2I}}}&{  {\mathbf{F}_{2R}}}\\
        {{-\mathbf{F}_{2R}}}&{{\mathbf{F}_{2I}}}
        \end{array}} \right)=\mathbf{AQP},
\end{eqnarray*}
\noindent where $\mathbf{A}=(\mathbf{F}_{1R}^T,\mathbf{F}_{1I}^T) \in \mathbb{C}^{N\times 2M}$,
\vspace{0.1cm}
\begin{eqnarray*}
\mathbf{P}=\left( {\begin{array}{cc}
        {{\mathbf{F}_{2R},}}&{ - {\mathbf{F}_{2I}}}\\
        {{\mathbf{F}_{2I},}}&{{\mathbf{F}_{2R}}}
        \end{array}} \right),
\mathbf{Q}=\left( {\begin{array}{cc}
        {{\mathbf{0}}}&{  {\mathbf{I}_{M}}}\\
        {{-\mathbf{I}_{M}}}&{{\mathbf{0}}}
        \end{array}} \right)
\end{eqnarray*}
\noindent are both $2M\times 2N$ matrices. Under the assumption that the number of RIS elements $N$ is large ($N\gg M$), the ranks of the three matrices $\mathbf{A}, \mathbf{P}$ and $\mathbf{Q}$ are at most $2M$ (full rank). Let
\begin{eqnarray*}
\mathbf{W}_1=\left( {\begin{array}{cc}
        \mathbf{B}_{R} & -\mathbf{B}_{I}\\
\mathbf{B}_{I} & \mathbf{B}_{R}
        \end{array}} \right).
\end{eqnarray*}

\noindent Thus we can deduce that $\textrm{rank}(\mathbf{W}_1)\leq\textrm{rank}(\mathbf{P})\leq 2M$, due to the fact that
\begin{eqnarray*}
\mathbf{W}_{1}=\left( {\begin{array}{cc}
        {\mathbf{AP}}\\
        {\mathbf{AQP}}
        \end{array}} \right)
=\left( {\begin{array}{cc}
        {\mathbf{A}}\\
        {\mathbf{AQ}}
        \end{array}} \right)\mathbf{P}.
\end{eqnarray*}

\vspace{0.1cm}
Therefore $\textrm{rank}(\mathbf{W})\leq 2M+2$. As long as $N\gg M$, $\mathbf{W}$ is a rank deficient matrix.
$\Box$

\subsection{Proof of Theorem \ref{theo:4}} \label{ap:ap3}
Since the feasibility violation reduces in each iteration, it must approach $0$ in the end as long as problem (\ref{eq:eq4}) is feasible. Thus it holds that
\begin{eqnarray} \label{eq:fv1}
\mathbf{y}^*=\mathbf{x}^*.
\end{eqnarray}
$\mathbf{x}^*$ and $\mathbf{y}^*$ must be the optimal solutions of the subproblems (\ref{eq:eq6}) and (\ref{eq:eq24}), respectively, because the penalty parameters satisfy (\ref{eq:eq76}).

\noindent 1. $\mathbf{x}^*$ satisfies the KKT conditions of the subproblem (\ref{eq:eq6}) for $\mathbf{x}$:
\begin{subequations}
\begin{eqnarray}
2\mathbf{A}_1\mathbf{x}^*+2\mathbf{a}_1+\nu\frac{\partial g(\mathbf{x})}{\partial \mathbf{x}}_{|\mathbf{x}=\mathbf{x}^*}=\mathbf{0},\label{eq:eq22.1}\\
(\mathbf{d}_{1}+\mathbf{F}_{1}\mathbf{x}^*)^{H}(\mathbf{d}_{2}+\mathbf{F}_{2}\mathbf{x}^*)=0.\label{eq:eq22.2}
\end{eqnarray}
\end{subequations}
where $\nu$ is the corresponding Lagrange multiplier, and $\mathbf{A}_1$ and $\mathbf{a}_1$ are defined below (\ref{eq:eq6}). Taking $\mathbf{A}_1$ and $\mathbf{a}_1$ into (\ref{eq:eq22.1}), we have
\begin{eqnarray}\label{eq:KKT1x}
-2\lambda\mathbf{F}_{1}^{H}(\mathbf{F}_{1}\mathbf{x}^*+\mathbf{d}_{1})
-2(1-\lambda)\mathbf{F}_{2}^{H}(\mathbf{F}_{2}\mathbf{x}^*+\mathbf{d}_{2})
+\boldsymbol{\mu}^*+\nu\frac{\partial g(\mathbf{x})}{\partial \mathbf{x}}_{|\mathbf{x}=\mathbf{x}^*}
=\mathbf{0}.
\end{eqnarray}

\noindent 2. $\mathbf{y}^*$ is the optimal solution of subproblem (\ref{eq:eq24}). The corresponding KKT conditions are:
\begin{eqnarray*}
\rho^*\mathbf{y}^*-(\rho^*\mathbf{x}^*+\boldsymbol{\mu}^*)+\boldsymbol{\tau }\odot\mathbf{y}^*=\mathbf{0},\label{eq:eq11.1}\\
\tau_l(|y_l^*|^2-1)=0, \tau_l\geq0, |y_l^*|^2\leq1, l=1,\ldots,N.\label{eq:eq11.2}
\end{eqnarray*}
where $\boldsymbol{\tau}=(\tau_1,\tau_2,\ldots,\tau_N)^T$, and $\tau_l$, $l=1,\ldots,N$ are the Lagrange multipliers. From (\ref{eq:fv1}) we have
\begin{subequations}
\begin{eqnarray}
\boldsymbol{\mu}^*=\boldsymbol{\tau}\odot\mathbf{x}^*\label{eq:KKT1y.1}\\
\tau_l(|x_l^*|^2-1)=0, \tau_l\geq0, |x_l^*|^2\leq1, l=1,\ldots,N.\label{eq:KKT1y.2}
\end{eqnarray}
\end{subequations}

Taking (\ref{eq:KKT1y.1}) into (\ref{eq:KKT1x}), we have
\begin{eqnarray}\label{eq:KKT1ap}
-2\lambda\mathbf{F}_{1}^{H}(\mathbf{F}_{1}\mathbf{x}^*+\mathbf{d}_{1})
-2(1-\lambda)\mathbf{F}_{2}^{H}(\mathbf{F}_{2}\mathbf{x}^*+\mathbf{d}_{2})
+\boldsymbol{\tau}\odot\mathbf{x}^*+\nu\frac{\partial g(\mathbf{x})}{\partial \mathbf{x}}_{|\mathbf{x}=\mathbf{x}^*}
=\mathbf{0}.
\end{eqnarray}
Together with (\ref{eq:eq22.2}) and (\ref{eq:KKT1y.2}), the KKT system of problem (\ref{eq:eq4}) is constructed. Thus $\mathbf{x}^*$ is the KKT point of problem (\ref{eq:eq4}).
$\Box$

\end{document}